\documentclass[prl, letterpaper, amsmath, amssymb, preprintnumbers, showpacs, superscriptaddress, twocolumn, floatfix]{revtex4-1}
\usepackage{graphicx}
\usepackage[usenames]{color}

\renewcommand{\eqref}[1]{(\ref{eq:#1})}

\begin{document}
\title{On the transverse M5-branes in matrix theory}

\author{Yuhma Asano}
\affiliation{
School of Theoretical Physics, Dublin Institute for Advanced Studies,
10 Burlington Road, Dublin 4, Ireland}
\author{Goro Ishiki}
\affiliation{Center for Integrated Research in Fundamental Science and Engineering (CiRfSE), University of Tsukuba, Tsukuba, Ibaraki 305-8571, Japan}
\affiliation{Graduate School of Pure and Applied Sciences, 
University of Tsukuba, Tsukuba, Ibaraki 305-8571, Japan}
\author{Shinji Shimasaki}
\affiliation{Research and Education Center for Natural Sciences, Keio University, Hiyoshi 4-1-1, Yokohama, Kanagawa 223-8521, Japan}
\author{Seiji Terashima}
\affiliation{
Yukawa Institute for Theoretical Physics, Kyoto 
University, Kyoto 606-8502, Japan}

\date{25 Jan 2017}

\preprint{DIAS-STP-16-14, UTHEP-701, YITP-16-149 }
% ------------------------------------------------------------------

% ------------------------------------------------------------------
\begin{abstract}
It has been a long-standing problem how 
the transverse M5-branes are described 
in the matrix-model formulations of M-theory.
We consider this problem for M-theory on 
the maximally supersymmetric pp-wave geometry,
which admits transverse spherical M5-branes with 
zero light-cone energy. 
By using the localization, 
we directly analyze the strong coupling region of the corresponding 
matrix theory called the plane wave matrix model (PWMM).
Under the assumption that the low energy modes of 
the scalar fields in PWMM become mutually 
commuting in the strong coupling region,
we show that the eigenvalue density of the $SO(6)$ scalars
in the low energy region exactly agrees with the shape of the 
spherical M5-branes in the decoupling limit. 
This result gives a strong evidence that the transverse 
M5-branes are indeed contained in the matrix theory and the theory 
realizes a second quantization of the M-theory.

\end{abstract}


\maketitle
% ------------------------------------------------------------------

%\section{intro} ------------------------------------------------------------
{\bf \textit{Introduction:}}
A nonperturbative formulation of M-theory in the light-cone frame is 
conjectured to be given by the matrix theory \cite{Banks:1996vh}.
The matrix theory is expected to achieve 
second quantization of M-theory, in which
all fundamental objects in M-theory are
described in terms of the internal degrees of freedom of matrices.
It has been shown that there exist matrix configurations 
corresponding to various objects in M-theory such as supergravitons, 
M2-branes and longitudinal M5-branes \cite{Banks:1996vh, deWit:1988wri, Banks:1996nn, Castelino:1997rv}.

On the other hand, the description of transverse M5-branes 
has not been fully understood. 
The charge of the transverse M5-branes is known to be 
absent in the supersymmetry algebra of the matrix theory, and hence
it seems to be impossible to construct matrix configurations 
for transverse M5-branes with nonvanishing charges. 

The absence of the M5-brane charge, however, does not prohibit the presence of 
M5-branes with compact world-volume, which have zero net charge.
It should be clarified whether such compact transverse M5-branes are 
included in the matrix theory.

The plane wave matrix model (PWMM)  
provides a very nice arena to understand this problem.
PWMM is the matrix theory for
M-theory on the maximally supersymmetric pp-wave background of the 11-dimensional supergravity \cite{Berenstein:2002jq}.
On this background, M-theory 
admits a stable spherical transverse M5-brane with vanishing light-cone energy.
In general, objects with zero light-cone energy in M-theory are mapped to 
vacuum states in the matrix theory. 
%Hence, the spherical transverse M5-brane on the pp-wave background 
%is expected to be realized 
%as a certain vacuum state in PWMM.
Thus, in finding the description of the spherical M5-branes, 
the target is restricted to the vacuum sector 
of PWMM.

As we will see below, vacua of PWMM are given by fuzzy sphere and are 
labeled by the partition of $N$, where $N$ is the matrix size of PWMM.
For each vacuum, the corresponding object with vanishing light-cone energy 
in M-theory was conjectured in \cite{Maldacena:2002rb}.
In particular, vacua corresponding to the spherical 
transverse M5-brane and its multiple generalization 
were specified. This conjecture was tested for the case of a single 
M5-brane by comparing the BPS protected
mass spectra of PWMM and those of the M5-brane.

In this Letter, we explicitly show that the spherical M5-brane emerges in the strong coupling regime of PWMM as the eigenvalue density of the $SO(6)$ scalars.
We apply the localization method \cite{Pestun:2007rz}
to PWMM and reduce the partition function to a simple matrix integral.
By evaluating the matrix integral in the strong coupling limit, we find that 
the eigenvalue density of the $SO(6)$ scalar matrices
forms a five-dimensional spherical shell and its 
radius exactly agrees with that of the spherical 
M5-brane in the M-theory on the pp-wave background.

%\section{M5 on pp-wave}-----------------------------------------------------
{\bf \textit{Spherical M5-brane on the pp-wave background:}}
We first review the spherical transverse M5-brane on the pp-wave background. 
The maximally supersymmetric pp-wave solution of 11-dimensional supergravity is given by
\begin{align}
ds^2 &=g_{\mu \nu}dx^\mu dx^\nu 
= -2dx^+ dx^- + \sum_{A=1}^9 dx^A dx^A \nonumber\\ 
& -\left( 
\frac{\mu^2 }{9}\sum_{i=1}^3x^ix^i +
\frac{\mu^2 }{36}\sum_{a=4}^9x^ax^a
\right) dx^+dx^+,
\nonumber\\
F_{123+} &= \mu,
\label{pp bg}
\end{align}
where $\mu$ is a constant parameter corresponding to 
the flux of the three form field. 
Throughout this Letter, we use the notation such that
$\mu, \nu =+,-,1,2,\cdots,9$, $A,B=1,2,\cdots,9$,
$i,j=1,2,3$ and $a,b=4,5,\cdots,9$.

We consider a single M5-brane in this background \cite{Maldacena:2002rb}.
Let $X^\mu(\sigma)$ be embedding functions of the M5-brane, 
where $\sigma^\alpha (\alpha=0,1,\cdots,5)$ are world-volume coordinates on the M5-brane.
The bosonic part of the M5-brane action is given by 
\begin{align}
S_{\rm M5}= -T_{\rm M5} \int d^6 \sigma \sqrt{-{\rm det}h_{\alpha \beta}}
+ T_{\rm M5} \int C_6,
\label{action M5}
\end{align}
where $h_{\alpha\beta}$ is the induced metric $h_{\alpha \beta}=g_{\mu\nu }(X)
\partial_{\alpha }X^\mu
\partial_{\beta }X^\nu$
and $C_6$ is the potential for the magnetic flux $dC_6 = *F_4$.
$T_{\rm M5}$ is the M5-brane tension, which is written in terms of 
the 11-dimensional Planck length $l_p$ as 
$T_{\rm M5} = \frac{1}{(2\pi)^5 l_p^6}$.
%As usually done in the literature, we rewrite the action into the Polyakov 
%type
%\begin{align}
%S_{\rm M5}&=-\frac{T_{M5}}{2}
%\int d^6 \sigma 
%\sqrt{-\gamma}
%\left(
%\gamma^{\alpha \beta}
%g_{\mu \nu}(X)\partial_{\alpha }X^\mu
%\partial_{\beta }X^\nu
%-1 \right) \nonumber \\
%&\quad + T_{\rm M5} \int C_6,
%\end{align}
%where $\gamma_{\alpha\beta}(\sigma)$ is an auxiliary world-volume metric 
%field.
%It is apparent that this action is invariant under the diffeomorphism 
%transformation
%for the world-volume coordinate. 
%% We fix the gauge by imposing
%% $\gamma_{0a}=0, \;\;\; \gamma_{00}= -\frac{4}{\nu^2}{\rm det}h_{ab}$, where $a,b=1,\cdots,5$ 
%% and $\nu$ is a constant which will be fixed by the light-cone momentum of the% M5-brane below.
%We fix this redundancy by imposing the light-cone gauge, where the 
%world-volume coodinate $\sigma^0$ is set
%as $X^+(\sigma)=\sigma^0$. Here, $X^\pm(\sigma)=(X^0\pm X^{10})/\sqrt{2}$.

By applying the standard procedure 
of the gauge fixing in the light-cone frame
\cite{Taylor:2001vb}, we obtain the light-cone 
Hamiltonian of the M5-brane as
\begin{align}
H_{\rm M5}
= & \int d^5 \sigma
\left[
\frac{V_5}{2p^+}
\left( P_A^2 +
\frac{T_{\rm M5}^2}{5!}
\{X^{A_1}, \cdots, X^{A_5} \}^2
\right)
\right.
\nonumber\\
& + 
\frac{p^+}{2V_5}
\left(
\frac{\mu^2}{9}(X^i)^2
+
\frac{\mu^2}{36}(X^a)^2
\right) \nonumber\\
&-\frac{\mu T_{\rm M5}}{6!}
\epsilon_{a_1a_2\cdots a_6}X^{a_1} 
\{X^{a_2},\cdots, X^{a_6} \}
\biggr],
\label{lc ham for M5}
\end{align}
where $V_5 = \pi^3$, $p^+$ is the total light-cone momentum, 
$P^A$ are 
the conjugate momenta of the transverse modes $X^A$
and the curly bracket is defined by
$\{f_1 ,\cdots, f_5 \} = \epsilon^{a_1\cdots a_5}
(\partial_{a_1} f_1)\cdots (\partial_{a_5}f_5)$.
%\begin{align}
%\{f_1 ,\cdots, f_5 \} = \epsilon^{a_1\cdots a_5}
%(\partial_{a_1} f_1)\cdots (\partial_{a_5}f_5).
%\end{align}

By noticing that the potential term for $X^a$ in (\ref{lc ham for M5}) can be 
rewritten as a perfect square,
%\begin{align}
%\frac{p^+\mu^2}{72 V_5}
%\left(
%X_{a_1} -\frac{6V_5T_{\rm M5}}{5!\mu p^+}
%\epsilon_{a_1a_2 \cdots a_6}
%\{X^{a_2},\cdots, X^{a_6} \}\right)^2.
%\end{align}
one can easily find the vacuum configuration as 
\begin{align}
P^A=0, \quad X^i=0, \quad X^{a}= r_{\rm M5}x^{a},
\label{spherical M5}
\end{align}
where $x^a$ are the embedding functions of the unit 5-sphere into $R^6$
satisfying
%\begin{align}
$x^ax^a=1, \; \{x^{a_1},\cdots, x^{a_5}\}=
\epsilon^{a_1a_2 \cdots a_6}x_{a_6}$.
%\end{align}
The constant $r_{\rm M5}$ is determined as
\begin{align}
r_{\rm M5}= \left( \frac{\mu p^+}{6\pi^3 T_{\rm M5}} \right)^{1/4}.
\label{radius of M5}
\end{align}
Thus, we find that 
the zero energy configuration is a spherical M5-brane 
with the radius given by (\ref{radius of M5}).

%\section{PWMM}-------------------------------------------------------------
{\bf \textit{Plane wave matrix model:}} 
The action of PWMM is obtained by 
the matrix regularization of a single M2-brane action on 
the pp-wave background \cite{Berenstein:2002jq}, which is given by the 
1+2 dimensional analogue of (\ref{action M5}).
The bosonic part of the action of PWMM is given by
\begin{align}
S=\frac{1}{g^2}
\int dt 
{\rm Tr}
\biggl[
&\frac{1}{2}\left(\frac{d}{dt}Y^A \right)^2 
-2 Y_i^2 -\frac{1}{2}Y_a^2 \nonumber\\
& \hspace{-1cm} +\frac{1}{4}[Y^A,Y^B]^2
-i\epsilon_{ijk} Y^i[Y^j,Y^k]
\biggr].
\label{action of PWMM}
\end{align}
In obtaining this action, we first apply the matrix regularization, where
the embedding functions $X^A(\sigma)$ of the M2-brane are mapped to 
$N\times N$ Hermitian matrices $Y^A$ as
\begin{align}
X^A(\sigma^0, \sigma^1, \sigma^2) 
\rightarrow \frac{\mu p^+}{12 \pi N T_{\rm M2}} Y^A (t).
\label{rescaling}
\end{align}
Here, $T_{\rm M2}=\frac{1}{(2\pi)^2l_p^3}$ is the tension of the M2-brane.
Poisson brackets and integrals on the spatial worldvolume 
are mapped to commutators and traces of matrices, respectively 
\cite{footnote1}.
The complicated factor in (\ref{rescaling}) is chosen so that the 
action (\ref{action of PWMM}) takes the simple form. 
In equation (\ref{rescaling}), the time coordinate $t$ is related to 
$\sigma^0$ by the same rescaling factor.
The coupling constant $g^2$ in (\ref{action of PWMM})
is related to the original  parameters in the M-theory by
\begin{align}
g^2 = \frac{T_{\rm M2}^2}{2\pi } \left( \frac{12 \pi N}{\mu p^+} \right)^3.
\label{redefining coupling}
\end{align}
%The covariant derivative $D$ in (\ref{action of PWMM}) is defined by
%$DY^A = \partial_t Y^A -i[A, Y^A]$.
%This is introduced to take into account the constraint of the form 
%$\{P^A, X^A\}=0$,
%which arises in the gauge-fixing in the light-cone frame
%\cite{Taylor:2001vb}.
%In the $A=0$ gauge, the Gauss law constraint 
%correctly reproduces this constraint.

Noticing that the potential for $Y^i$ 
in (\ref{action of PWMM}) forms a perfect square,
one can easily find the vacuum configuration of PWMM as
\begin{align}
Y^i=2L^i, \;\; Y^a=0.
%\;\; A=0.
%X^i = \frac{\mu p^+}{6\pi N T_{\rm M2}} L^i　
\label{fuzzy sphere vacuum}
\end{align}
Here, $L^i$ are 
$N$-dimensional representation matrices of the $SU(2)$ generators. 
The representation can be 
reducible and one can make an irreducible decomposition,
\begin{align}
L_i = \bigoplus_{s=1}^\Lambda L_i^{[n_s]},
%\bigoplus_{s=1}^\Lambda L_i^{[n_s]} \otimes 1_{N_s},
\label{irre decom}
\end{align}
where $L_i^{[n]}$ stand for the generators in the 
$n$-dimensional irreducible representation and $\sum_{s=1}^{\Lambda}n_s =N $. Thus, the vacua are labeled 
by the partition of $N$, $\{n_s | n_s \ge n_{s+1}, \sum_s n_s = N \}$.

% \section{The conjecture on the spherical M5-brane}-----------------------------
{\bf \textit{The conjecture on the spherical M5-brane:}} 
The vacuum of the form (\ref{fuzzy sphere vacuum}) is the fuzzy sphere
configuration and hence it has a clear interpretation as 
a set of spherical M2-branes. 
Indeed, the M-theory on the pp-wave background allows
zero energy spherical M2-branes as well. The commutative limit of the 
fuzzy sphere (\ref{fuzzy sphere vacuum}), where $n_s$ become large, 
can naturally be identified with those M2-branes in M-theory.

On the other hand, the correspondence between 
(\ref{irre decom}) and the spherical M5-brane can be understood 
by introducing a dual way of looking at the Young tableau of the vacuum.
For the vacuum (\ref{irre decom}), let us consider the Young tableau
corresponding to the partition $\{n_s \}$
such that the length of the $s$th column is given by $n_s$. 
Let us denote by $m_k$ the length of the $k$th row, where $k$ runs from 1 to 
${\rm max}\{n_s\}$.
It was conjectured in \cite{Maldacena:2002rb} 
that when $m_k$ are large, the vacuum corresponds to 
multiple M5-branes, where the number of M5-branes is given by 
$N_5:={\rm max}\{n_s\}$ and each M5-brane carries the light-cone momentum 
proportional to $m_k (k=1,2, \cdots, N_5)$.

%This conjecture is highly nontrivial. For example, let us consider 
%the case with $N_5=1$, which is just the trivial vacuum $Y^i=0$ of PWMM.
%The above conjecture claims that even this trivial vacuum corresponds to 
%the single spherical M5-brane in M-theory.
%The trivial vacuum configuration 
%Note that This is possible since M-theory is supposed to be realized in 
%a strong coupling regime of PWMM, where the typical configuration of matrices 
%is nontrivial.

In what follows, we test this conjecture focusing on the vacua 
such that the partition is of the form,
\begin{align}
L_i = L_i^{[N_5]} \otimes 1_{N_2}.
\label{specific vacuum}
\end{align}
$N_2$ and $N_5$ satisfy $N_2N_5=N$ and 
correspond to the number of M2- and M5- branes, respectively.
In order to describe the M5-brane, we consider the limit, 
\begin{align}
N_2 \rightarrow \infty,  \quad N_5 \; : \; \text{fixed}.
\label{M5 limit}
\end{align}
With the above interpretation, this limit corresponds to 
$N_5$ M5-branes, each of which carries the light-cone momentum with an
equal amount.

In order to isolate the degrees of freedom of the M5-branes in PWMM,  
the 't Hooft coupling of PWMM should also be sent to infinity in taking 
the limit (\ref{M5 limit}) \cite{Maldacena:2002rb}.
This can be understood as follows.
We first rewrite the metric 
(\ref{pp bg}) so that the compactified direction $x^-$ is
orthogonal to the other directions as
\begin{align}
ds^2 
%&= -2 dx^+ dx^-  -\frac{\mu^2 r^2}{36} dx^+dx^+ + r^2 d\Omega_2^5 +\cdots
%\nonumber\\
&= -\frac{\mu^2 r^2 }{36}d\tilde{x}^+ d\tilde{x}^+ 
+\frac{36}{\mu^2 r^2 }d\tilde{x}^- d\tilde{x}^- + r^2 d\Omega_2^5 + \cdots ,
\label{rescaled metric}
\end{align}
where $r^2=\sqrt{x^ax^a}$ is the radius of the 5-sphere and 
$\tilde{x}^+ = x^+ -\frac{36}{\mu^2 r^2}x^-$, $\tilde x^-=x^-$.
The physical compactification radius is then given by
$\tilde R\sim R/(\mu r)$, where $R$ is the original compactification radius of 
M-theory.
Upon compactification, transverse M5-branes 
in the M-theory become NS5-branes in the type IIA superstring theory.
The world-volume theory of the NS5-branes is known as the little string theory, which has a characteristic scale given by the string tension $\sim l_s^{-2}$.
For the spherical NS5-brane with the radius 
%$r_{\mathrm{M5}}$ in
(\ref{radius of M5}), 
the theory is controlled by the dimensionless combination $r^2_{\rm M5}/l_s^2$.
We keep this ratio finite to obtain an interacting theory on the NS5-branes 
while we send $r_{\rm M5}$ to infinity to make the bulk gravity decouple.
By using (\ref{radius of M5}) and the well known relation $l_s\sim(l_p^3/\tilde R)^{1/2}$, 
we find that the decoupling limit of the NS5-brane is given by
$p^+\to \infty$ with $R^4p^+$ fixed \cite{footnote2}.
The M5-brane in 11 dimensions is recovered by further 
taking $R^4p^+$ to be large.
Finally, by combining this observation with (\ref{M5 limit}),
we find that the decoupling limit of the M5-brane 
is written in terms of the parameters of PWMM as
\begin{align}
N_2\to \infty, \quad N_5 \; : \; \text{fixed}, \quad 
\lambda \to \infty, \quad \frac{\lambda}{N_2} \rightarrow 0,
\label{decoupling limit of M5}
\end{align}
where $\lambda=g^2N_2$ is the 't Hooft coupling of PWMM.
Thus, the decoupling limit of the M5-brane corresponds to the strong coupling limit in the 't Hooft limit.

%\section{M5 from PWMM} ------------------------------------------------------
{\bf \textit{Spherical M5-branes from PWMM:}} 
Let us analyze PWMM in the decoupling limit of the M5-brane
by using the localization method.
We consider the following scalar field:
\begin{align}
\phi (t) = Y_3(t) + i(Y_8(t) \sin ( t) + Y_9(t) \cos (t)).
\label{def of phi}
\end{align}
This field preserves 1/4 of the whole supersymmetries in PWMM and 
any expectation values made of only $\phi $ can be computed by the 
localization method \cite{Pestun:2007rz}.
%(Precisely speaking, the localization computation is performed for the theory 
% obtained by the double Wick-rotation for the time and $Y_9(t)$ directions.
%Then, $\phi(t)$ is shown to be invariant under 4 supercharges.)
The computation is done in the Euclidean theory, which is 
obtained by performing the Wick rotaton $t \rightarrow -i\tau$.
Since we are interested in PWMM around a specific vacuum
(\ref{specific vacuum}),
we impose the boundary conditions such that all the fields take the vacuum 
configurations at $\tau \to \pm \infty$.
With this boundary condition, 
the localization computation leads to the following equality
\cite{Asano:2012zt,Asano:2014vba,Asano:2014eca}:
\begin{align}
\langle \prod_I {\rm Tr}f_I(\phi (t_I)) \rangle
= 
\langle \prod_I {\rm Tr}f_I(2L_3+iM) \rangle_{MM},
\label{result of localization}
\end{align}
where $f_I$ are arbitrary smooth functions of $\phi$,
$2L_3$ is the vacuum configuration for $Y_3$, and
$M$ is an $N\times N$ Hermitian matrix which commutes with all $L_a (a=1,2,3)$.
For the vacuum given by (\ref{specific vacuum}), $M$ takes the form, 
$M= {\bf 1}_{N_5} \otimes  \tilde{M}$,
where $\tilde{M}$ is an $N_2\times N_2 $ Hermitian matrix.
The expectation value $\langle \cdots \rangle$ in the left-hand side of 
(\ref{result of localization}) 
is taken with respect to the original partition function of PWMM around (\ref{specific vacuum}),
while that in the right-hand side, $\langle \cdots \rangle_{MM}$,
is taken with respect to the matrix integral,
\begin{align}
Z&= \int \prod_i dq_i e^{-\frac{2N_5}{g^2}\sum_{i}q_i^2} 
\prod_{J=0}^{N_5-1} \prod_{j=1}^{N_2-1}\prod_{i=j+1}^{N_2}
\nonumber \\
&\times
\frac{\{(2J+2)^2+(q_{i}-q_{j})^2\} \{(2J)^2 +(q_{i}-q_{j})^2 \} }
{\{(2J+1)^2 +(q_{i}-q_{j})^2\}^2}.
\label{one matrix}
\end{align}
Here, $q_i$ $(i=1,2,\cdots, N_2)$ are the eigenvalues of $\tilde{M}$ 
\cite{footnote3}.

In the decoupling limit (\ref{decoupling limit of M5}),
the saddle point approximation becomes exact in evaluating the matrix integral.
We first introduce the eigenvalue density for $q_i$ by
$\rho(q)= \frac{1}{N_2}\sum_{i=1}^{N_2}\delta (q-q_i)$,
%\begin{align}
%\rho(q)= \frac{1}{N_2}\sum_{i=1}^{N_2}\delta (q-q_i),
%\label{eigenvalue distribution def}
%\end{align}
which is normalized as $\int_{-q_m}^{q_m}dq \rho(q) =1$.
Here, we assume that $\rho(q)$ has a finite support $[-q_m,q_m]$.
In the large-$\lambda$ limit, the saddle point equation of the 
partition function (\ref{one matrix}) is reduced to
\begin{align}
\beta = \pi \rho(q) + \frac{2N_5}{\lambda }q^2 
- \int_{-q_m}^{q_m} dq' \frac{2N_5}{(2N_5)^2+(q-q')^2}
\rho (q'),
\label{spe}
\end{align}
where $\beta$ is the Lagrange multiplier for the normalization of $\rho$ and 
we used the fact that $q_m/N_5\gg 1$ in this limit. 
The solution of (\ref{spe}) is given by
\begin{align}
\rho (q) = \frac{8^{\frac{3}{4}}}{3\pi \lambda^{\frac{1}{4}}}
\left[1-
\frac{q^2}{q_m^2}
\right]^{\frac{3}{2}}, \;\;\;  q_m=(8\lambda )^{\frac{1}{4}}, 
\;\;\; \beta = \frac{8^{\frac{1}{2}}N_5}{\lambda^{\frac{1}{2}}}.
\label{density solution}
\end{align}

By using (\ref{result of localization}) and the solution for the 
eigenvalue density (\ref{density solution}), 
we can compute any operator made of $\phi $.
In particular, let us consider the resolvent of $\phi $ defined by
${\rm Tr}(z-\phi)^{-1}$. 
According to the result of the localization 
(\ref{result of localization}), the expectation value 
of this operator is equal to 
that of ${\rm Tr}(z-2L_3-iM)^{-1}$ in the matrix 
integral (\ref{one matrix}).
Note that the support of the eigenvalue density of $M$ is much 
larger than that of $L_3$ in the decoupling limit. 
Thus, this shows that, with the suitable normalization as in 
(\ref{rescaling}), 
the spectrum of $\phi$ lies on the imaginary axis and is 
given by $\rho(q)$ in (\ref{density solution}) in the decoupling limit. 

One might expect that the density $\rho(q)$ can
be identified with the eigenvalue density of one of the $SO(6)$ scalars. 
However, such identification would lead to a contradiction with the result 
in \cite{Polchinski:1999br}, where the typical scale of the 
distribution of the scalar fields is shown to be of order of $\lambda^{1/3}$ 
\cite{Maldacena}. 
A consistent identification can be made by 
considering the low energy region 
in the discussion in \cite{Polchinski:1999br}.
This will be achieved by taking a time average of operators.
Note that the equation (\ref{result of localization}) is time-independent
\cite{Asano:2014vba}. We integrate the both sides over a 
very short time interval with length $1/C$, where $C$ is a constant
much smaller than the typical scale $\lambda^{1/3}$ but 
much larger than the relevant scale for (\ref{density solution}). 
This time average projects the operators to the low energy modes. 
Assuming that the low energy modes of the scalar fields $Y^A$ 
are mutually commuting in the strong coupling limit, we find that
%\begin{align}
%\frac{1}{N}\langle {\rm Tr}(\bar{Y}^9)^n (0) \rangle = 
%\int_{-q_m}^{q_m}dq \rho(q) q^n,
%\label{main result}
%\end{align}
%where $\bar{Y}^9= \int^{t_0}_0 Y_()$
$\rho(q)$ is identified with the eigenvalue density of the low energy modes 
of $Y^a$. 
Here, the commutativity of the low energy modes are 
naturally expected since the matrix model is expected to describe the classical geometric objects. See also \cite{Berenstein:2008eg} for the commutativity in 
the strong coupling limit.

With the assumption of the commutativity
as well as the $SO(6)$ symmetry in PWMM, 
%which is not broken spontaneously, 
we then define the joint eigenvalue distribution $\tilde\rho$ 
of all the $SO(6)$ scalar fields.
We define $\tilde\rho$ as the $SO(6)$ symmetric uplift of $\rho$:
\begin{align}
\int d^6 x^a \tilde{\rho}(r) x_9^{2n}
= \left(\frac{\mu p^+}{12 \pi N T_{\rm M2}}\right)^{2n}
\int^{q_m}_{-q_m}dq \rho(q) q^{2n}
\label{uplift}
\end{align}
for any $n$.
Here, $r=\sqrt{\sum_{a}x_a^2}$ and $\tilde{\rho}$
is normalized as
$\int d^6x^a \tilde{\rho} (r)=1$.
Note that $\tilde{\rho}$ depends only on $r$
because of the $SO(6)$ symmetry. 
The first factor on the right-hand side of (\ref{uplift}) just reflects
the rescaling (\ref{rescaling}), so that $\tilde\rho(r)$ can be 
thought of as a density function in the original target space.

The unique solution to (\ref{uplift}) is given by a
spherical shell in $R^6$ as
\begin{align}
\tilde{\rho} (r)= \frac{1}{V_5r_0^5}\delta(r-r_0),\;\;\;\; 
r_0= \left( \frac{\mu p^+}{6\pi^3 N_5 T_{\rm M5}} \right)^{1/4}.
\label{spherical distribution}
\end{align}
%The distribution (\ref{result of localization}) is 
%just the projection of the spherical distribution 
%(\ref{spherical distribution}) onto a one-dimensional axis in $R^6$.
For $N_5=1$, the shape of the density function 
(\ref{spherical distribution}) exactly 
agrees with the shape of the spherical M5-brane on the pp-wave background. 
In particular, the radius $r_0$ agrees with (\ref{radius of M5}).
Thus, under the above assumption, this shows that the transverse M5-brane is 
formed by the eigenvalue density of the low energy modes of the 
$SO(6)$ scalars.

For $N_5>1$, $r_0$ in (\ref{spherical distribution}) is interpreted as the 
radius of the multiple spherical M5-branes. 
The $N_5$-dependence of $r_0$ coincides with the conjectured form 
in \cite{Maldacena:2002rb} 
based on an observation on perturbative expansions in PWMM.

%\section{summary} ----------------------------------------------------
{\bf \textit{Summary:}} 
In this Letter, we considered the matrix theoretical 
description of the spherical transverse M5-branes with vanishing light-cone 
energy in M-theory on the maximally supersymmetric pp-wave background.
Following the proposal in \cite{Maldacena:2002rb}, 
we considered PWMM expanded around the vacuum associated with the M5-branes.
We applied the localization to this theory and
obtained an eigenvalue integral. 
We then analyzed and solved the eigenvalue integral in the 
decoupling limit of the M5-branes, which corresponds to the 
strong coupling limit of PWMM.
Finally, under the assumption that the 
low energy modes of the scalar fields become mutually commuting 
in the strong coupling limit,
we found that the eigenvalue density of the
low energy modes of $SO(6)$ scalar fields
forms a 5-dimensional spherical shell and 
the radius of the spherical shell exactly agrees with 
that of the M5-brane in M-theory.
Thus, we concluded that the M5-brane in M-theory is formed by
the eigenvalue density of the $SO(6)$ scalar fields in the low energy region.
We also computed the radius of the multiple M5-branes.

% ------------------------------------------------------------------
{\bf \textit{Acknowledgments:}} 
We thank J. Maldacena and H. Shimada for valuable discussions.
The work of G. I. was supported, in part, 
by Program to Disseminate Tenure Tracking System, 
MEXT, Japan and by KAKENHI (16K17679).  
S. S. was supported by the MEXT-Supported Program for the Strategic Research Foundation at Private Universities
 “Topological Science” (Grant No. S1511006).


\begin{thebibliography}{99}
\bibitem{footnote2} 
Note that we are 
interested in the case where $l_p$ and $\mu$ are fixed.
\bibitem{footnote3}
In the derivation of (\ref{result of localization}), possible instanton 
corrections are ignored. However, these corrections are indeed negligible
in the decoupling limit (\ref{decoupling limit of M5}).
\end{thebibliography}
\end{document}